\documentclass[aps,twocolumn,superscriptaddress]{revtex4}

\usepackage{graphicx}
\usepackage{bm}

\begin{document}

\preprint{}

\title{Uniform mixing of antiferromagnetism and high-$T_{\rm c}$ superconductivity in multilayered copper oxides  
Ba$_2$Ca$_{n-1}$Cu$_n$O$_{2n}$F$_2$ with apical fluorines ($n=2,3,4$):
 $^{63}$Cu-NMR/NQR and $^{19}$F-NMR}

\author{S. Shimizu}
\email[]{e-mail: shimizu@nmr.mp.es.osaka-u.ac.jp}
\author{T. Sakaguchi}
\author{H. Mukuda}
\author{Y. Kitaoka}
\affiliation{Department of Materials Engineering Science, Osaka University, Osaka 560-8531, Japan }
\author{P. M. Shirage}
\author{Y. Kodama}
\author{A. Iyo}
\affiliation{National Institute of Advanced Industrial Science and Technology (AIST), Umezono, Tsukuba 305-8568, Japan}

\date{\today}

\begin{abstract}
We report $^{63}$Cu-NMR/NQR and $^{19}$F-NMR studies on the multilayered high-$T_{\rm c}$ copper oxides Ba$_2$Ca$_{n-1}$Cu$_n$O$_{2n}$F$_2$ with $n=2,3,4$, where $n$ is the number of CuO$_2$ planes.  It is revealed that bi-layered Ba$_2$CaCu$_2$O$_4$F$_2$ is an underdoped superconductor with {\it hole carriers}, which are introduced into CuO$_2$ planes by an unexpected deviation from the nominal content of apical fluorines. In a previous paper, we proposed a {\it self-doping} mechanism as the origin of carrier doping in $n=3$ and $n=4$; in the mechanism, electrons are transferred from the inner CuO$_2$ plane (IP) to the outer one (OP). However, since it has been found that the bi-layered compound is hole doped, we have reexamined the superconducting and magnetic properties in $n=3$ and $n=4$ by $^{63}$Cu-NMR/NQR and $^{19}$F-NMR. The extensive NMR studies have confirmed that the apical-fluorine compounds are not {\it self-doped} but {\it hole-doped}, and that antiferromagnetism (AFM) and superconductivity (SC) coexist in a single CuO$_2$ plane.
In $n=4$, the AFM ordering occurs at $T_{\rm N}$ = 80 K, well above $T_{\rm c}=55$ K, where the respective AFM moments are $M_{\rm AFM}=0.11$ $\mu_{\rm B}$ and 0.18 $\mu_{\rm B}$ at the OP and the IP. In $n=3$, on the other hand, the underdoped single IP exhibits a spontaneous moment $M_{\rm AFM}=0.12$ $\mu_{\rm B}$ at low temperatures and a peak in the nuclear-spin-lattice relaxation rate $1/T_1$ of $^{19}$F at $T_{\rm N}=23$ K, much lower than $T_{\rm c} =$ 76 K.
We note that the increase in the number of IPs from one to two results in the strengthening of the interlayer coupling; $T_{\rm N}$ increases as the interlayer coupling becomes stronger, although the doping levels for both compounds are comparable. 
Consequently, we conclude that the uniform mixing of AFM and SC is a general property inherent to a single CuO$_2$ plane in an underdoped regime for hole-doping. This conclusion incorporates the ARPES results on the n=4 compound [Y. Chen {\it et al.}, Phys. Rev. Lett. 97, 236401 (2006)]; it was found that the two Fermi sheets of the IP and OP are observed, and that the SC gap opens at the IP and OP below $T_{\rm c}$ = 55 K.

\end{abstract}

\pacs{74.72.Jt; 74.25.Ha; 74.25.Nf}

\maketitle

\section{Introduction}

\begin{figure}[h]
\begin{center}
\includegraphics[width=0.75\linewidth]{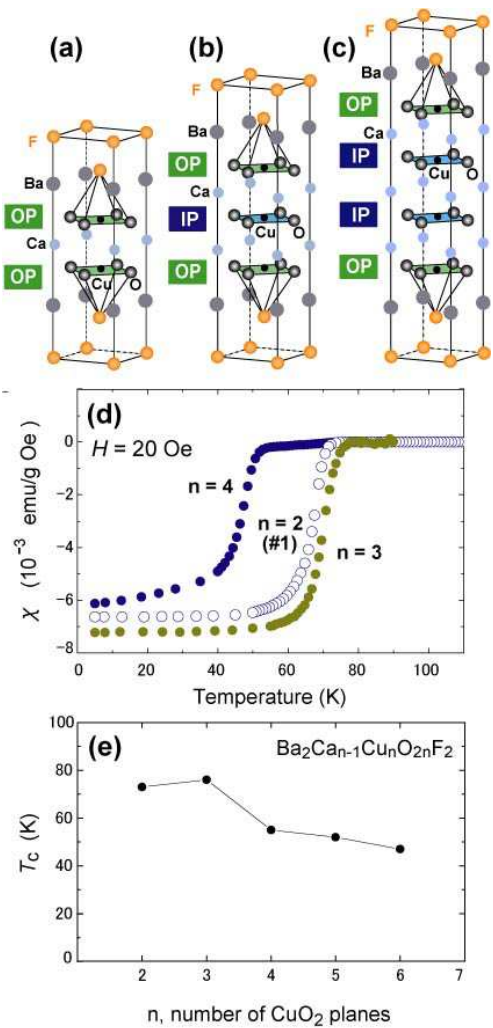}
\end{center}
\caption{\footnotesize (color online) Crystal structures of the (a) $n$ = 2, (b) $n$ = 3, and (c) $n=4$ compounds in the Ba$_2$Ca$_{n-1}$Cu$_n$O$_{2n}$F$_2$ series \cite{Iyo2}. The $n$ = 3 and $n=4$ compounds include two crystallographically different CuO$_2$ planes, namely, IP and OP, whereas the $n$ = 2 compound only includes OPs. (d) $T$ dependence of dc-susceptibility. $T_{\rm c}$ was determined as the temperature below which susceptibility begins to decrease rapidly. (e) Plots of $T_{\rm c}$  vs. the number of CuO$_2$ layers ($n$) \cite{Iyo_unpublished}.}
\label{fig:crystal}
\end{figure}

Despite more than 22 years of intensive research, an origin of high-temperature superconductivity (HTSC) has not been well understood yet. 
A strong relationship between antiferromagnetism (AFM) and superconductivity (SC) is believed to be a key to understand the origin of the remarkably high SC transition \cite{Anderson,Inaba,Zhang,Himeda,Sidis,Lake,TKLee,Demler,Lee,Miller,Moriya}. 
However, there is still no universally accepted theory for the mechanism of cuprate superconductors. 
Remarkably, the highest recorded $T_{\rm c}$ in cuprates was observed in a Hg-based three-layered HgBa$_2$Ca$_2$Cu$_3$O$_{y}$ (Hg-1223) \cite{Schilling}, which has multilayered structure of CuO$_2$ planes. Cuprates with three or more layers comprise inequivalent types of CuO$_2$ layers; one is an outer CuO$_2$ plane (OP) in a five-fold pyramidal coordination and the other, an inner plane (IP) in a four-fold square coordination (see Fig.\ref{fig:crystal}). Site-selective $^{63}$Cu-NMR studies on multilayered cuprates have revealed that IP has less hole carriers than OP does, and that they exhibit homogeneous hole-doping due to the flatness of the layers \cite{Tokunaga,Kotegawa2001}. NMR linewidths assure that the flatness of the layers in multilayered structures; linewidths for multilayered structures, less than 200 Oe, is several times smaller than those for LaSrCuO an YBaCuO \cite{Takigawa,Ohsugi}.
In particular, five-layered superconductors have an AFM order due to large charge imbalance \cite{Kotegawa2004,Mukuda2006}; in HgBa$_2$Ca$_4$Cu$_5$O$_{y}$ (Hg-1245) \cite{Kotegawa2004}, two optimally doped OPs are responsible for SC with $T_{\rm c}=108$ K, whereas three underdoped IPs are responsible for AFM with $T_{\rm N}=60$ K. On the basis of various five-layered cuprates, the novel phase diagram of doped CuO$_2$ planes has been presented as a function of doping; an AFM phase is robust for hole-doping of up to 16 $\sim$ 17 \% and it coexists with SC in a region where the SC order parameter begins to develop \cite{Mukuda2006, Mukuda2008}. $T_{\rm c}$ exhibits a maximum immediately outside a quantum critical point (QCP) at which the AFM order collapses, suggesting the intimate relationship between AFM and SC. 

Multilayered copper oxides Ba$_2$Ca$_{n-1}$Cu$_n$O$_{2n}$F$_2$ are known as a relatively-new family of multilayered cuprates \cite{Iyo2,Iyo1}, where $n$ is the number of CuO$_2$ planes. 
Figures \ref{fig:crystal}(a-c) show the crystal structures of $n$-layered cuprates with $n=2,3,4$, which are denoted as the $n$ compound in this paper. 
In an ideal nominal content of fluorine, the apical sites of the OP are fully occupied by F$^{-1}$ and hence the formal Cu valence is just +2. Eventually, it is expected that these compounds could act as Mott insulators; however, all of them exhibit SC. The occurrence of the SC in the $n=4$ compound has been attributed to {\it self-doping} in which charge carriers are transferred between the IP and the OP; this was first proposed based on ARPES (angle resolved photoemission spectroscopy) \cite{Chen} and a band calculation \cite{OK}. 
Our previous NMR study has supported the {\it self-doping} mechanism, assuming that the Cu valence is +2 on average; the formal copper valence Cu$^{+2}$ would be expected from the stoichiometry \cite{Shimizu}. Since then, a bi-layered Ba$_2$CaCu$_2$O$_4$F$_2$ (the $n$ = 2 compound(\#1)), which has only OP and no IP, has also been synthesized; it exhibits SC with $T_{\rm c}=$ 73 K \cite{Iyo_unpublished}. The {\it self-doping} mechanism can not apply to this compound because it has no IPs. Therefore, NMR studies on the $n=2$ compound(\#1) and a related, partially oxidized Ba$_2$CaCu$_2$O$_4$(F$_{1.6}$O$_{0.4}$) (the $n$ = 2 compound(\#2)) will provide us with an opportunity to obtain an understanding of whether the {\it self-doping} mechanism occurs in the $n=3$ and $n=4$ compounds.

In this paper, we report Cu-NMR/NQR studies on the $n=2$ compounds(\#1) and (\#2), which reveal that SC in the $n=2$ compound(\#1) occurs due to {\it hole carriers} doped in association with a possible replacement of O$^{-2}$ for F$^{-1}$.  We compare the results of the $n=2$ compounds with those of the $n=3$ and $n=4$ compounds. We find that hole carriers are unexpectedly introduced into both the OP and the IP by a deviation from the nominal compositions;
Cu$^{+2}$ realizes only in the case of the stoichiometry under total charge neutrality. As a result, we rule out the possibility of the {\it self-doping} mechanism in the $n=3$ and $n=4$ compounds. We present the following results. The $n=2$ compounds(\#1) and (\#2) are hole-doped superconductors with $T_{\rm c}$ = 73 K and 105 K, respectively. The $n=3$ compound is an antiferromagnetic superconductor with $T_{\rm c}$ = 76 K. The underdoped single IP shows the AFM order by an interlayer magnetic coupling through three layers, namely, SC-OP, BaF, and SC-OP (see Fig.\ref{fig:crystal}(b)).
The $n=4$ compound is also an antiferromagnetic superconductor. The two IPs trigger the AFM order with $T_{\rm N}$ = 80 K, whereas the two OPs show the SC with $T_{\rm c}=55$ K; interestingly, $M_{\rm AFM}$(OP) develops upon cooling and becomes 0.11 $\mu_{\rm B}$ at 1.5 K. Consequently, we conclude that the uniform mixing of AFM and SC is a general property inherent to a single CuO$_2$ plane in an underdoped regime for hole-doping. This conclusion incorporates the ARPES results on the n=4 compound \cite{Chen}; it was found that the two Fermi sheets of the IP and OP are observed, and that the SC gap opens at the IP and OP below $T_{\rm c}$ = 55 K.

\section{Experimental details}

Polycrystalline powder samples used in this study were prepared by a high-pressure synthesis technique, as described elsewhere \cite{Iyo1,Iyo2}. Powder X-ray diffraction measurements indicate that these compounds almost entirely comprise a single phase. Figure \ref{fig:crystal}(e) shows the variation of $T_{\rm c}$ for Ba$_2$Ca$_{n-1}$Cu$_n$O$_{2n}$F$_2$ as a function of $n$ \cite{Iyo_unpublished}. The $T_{\rm c}$ was determined by an onset of diamagnetism using a dc SQUID magnetometer for the powder samples at $H=$ 20 Oe, as shown in  Fig. \ref{fig:crystal}(d).
The characteristics of the samples used in this study, $n=$ 2 $-$ 4, are summarized in Table \ref{samples}.

For NMR measurements, the powder samples, which were aligned along the $c$-axis at an external field ($H$) of $\sim$ 15 T, were fixed using stycast 1266 epoxy. The NMR experiments were performed by the conventional spin-echo method in a temperature ($T$) range of 1.5 $-$ 325 K with $H$ perpendicular or parallel to the $c$-axis.
Generally, the Hamiltonian for a Cu nuclear spin with $I=3/2$ is described by the Zeeman interaction due to a magnetic field $H$ and the nuclear quadrupole interaction (NQI) as follows:
\begin{equation}
{\cal H}={\cal H_{\rm Z}}+{\cal H_{\rm Q}}=-\gamma_{\rm N}\hbar {\bm I} \cdot {\bm H}+\frac{eV_{\rm zz}Q}{4I(2I-1)}(3I_{\rm z}^2-I(I+1)),
\label{eq:hamiltonian}
\end{equation}
where $\gamma_{\rm N}$ is the Cu nuclear gyromagnetic ratio and $H$ is perpendicular to the $c$-axis. Note that the quadrupole frequency $\nu_{\rm Q}\equiv 3eQV_{\rm zz}/2h I(2I-1)$, where $Q$ is the nuclear quadrupole moment and 
$V_{\rm zz}$, the electric field gradient (EFG) along the $c$-axis at the Cu nuclear site. Analyzing NMR spectra makes it possible to distinguish local electronic states at two different Cu sites, OP and IP, by the differences in Knight shift ($K$), NQR frequency ($\nu_{\rm Q}$), internal field ($H_{\rm int}$) or AFM ordered moment ($M_{\rm AFM}$), and so on.

\begin{table}[htbp]
\caption[]{Samples used in the present NMR studies. $T_{\rm c}$s of the samples are determined by an onset of diamagnetism using a dc SQUID magnetometer. The hole densities per CuO$_2$ plane ($N_{\rm h}$) are evaluated from the Knight shift measurement (see text).}
\label{samples}
\begin{center}
  \begin{tabular}{cccc}
Sample  & Nominal composition & $T_{\rm c}$(K) & $N_{\rm h}$  \\
    \hline
$n=2$($\sharp$1)  & Ba$_2$CaCu$_2$O$_{4}$F$_2$                 &  73  & 0.174  \\
$n=2$($\sharp$2)  & Ba$_2$CaCu$_2$O$_{4}$(F$_{1.6}$O$_{0.4}$)  & 105  & 0.225   \\
$n=3$             & Ba$_2$Ca$_2$Cu$_3$O$_{6}$F$_2$             &  76  & $^*$0.160 (OP)  \\ 
$n=4$             & Ba$_2$Ca$_3$Cu$_4$O$_{8}$F$_2$             &  55  & $^*$0.148 (OP)  \\
    \hline
    \end{tabular}
 \end{center}
\footnotesize{$*$) In particular, the $N_{\rm h}$s of the OPs are shown for the $n=$ 3 and 4 compounds.}
\end{table}

\section{Results and Discussions}

\subsection{Cu-NMR/NQR studies on the $n$ = 2 compounds}
\begin{figure}[htpb]
\begin{center}
\includegraphics[width=0.75\linewidth]{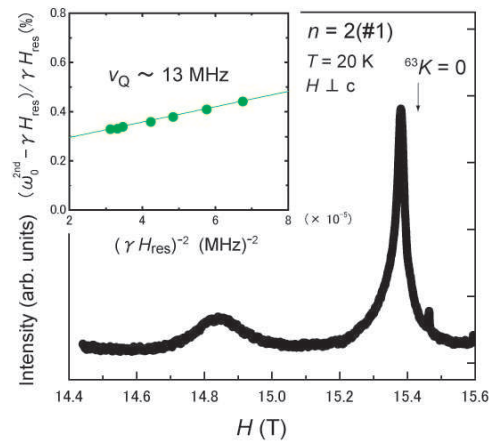}
\end{center}
\caption{\footnotesize (color online)  $^{63}$Cu-NMR spectrum for the $n=2$ compound($\sharp$1) at $T$ = 20 K with $H$ perpendicular to the $c$-axis. The right and left peaks correspond to a central peak ($-1/2 \Leftrightarrow$ 1/2 transition) and one of the satellite peaks ($-1/2 \Leftrightarrow$ $-3/2$ transition), respectively. 
The inset shows a plot of $(\omega_0^{2nd}-\gamma_{\rm N}H_{\rm res})/\gamma_{\rm N}H_{\rm res}$ as a function of $(1/\gamma_{\rm N}H_{\rm res})^2$ for the central peak. A slope in this plot yields $^{63}\nu_{\rm Q}$ (see text). }
\label{fig:spe2F2}
\end{figure}

Bi-layered Ba$_2$CaCu$_2$O$_4$F$_2$, which has two OPs (see Fig. \ref{fig:crystal}(a)), exhibits SC below $T_{\rm c}$ = 73 K. When all the apical sites of the OP are ideally occupied by F$^{-1}$, the formal Cu valence should be just +2. Therefore, a Mott insulating state should be expected as in undoped CuO$_2$ systems. However, off-stoichiometry, a deviation from the nominal content of F$^{-1}$, can introduce carriers into the OPs. We have investigated two kinds of bi-layered samples, the nominal compositions of which are Ba$_2$CaCu$_2$O$_4$F$_2$ (the $n$ = 2 compound(\#1)) and Ba$_2$CaCu$_2$O$_4$(F$_{1.6}$O$_{0.4}$) (the $n$ = 2 compound(\#2)), in order to understand the carrier-doping mechanism and the SC properties in the apical-fluorine system. 

\begin{figure}[tpb]
\begin{center}
\includegraphics[width=0.75\linewidth]{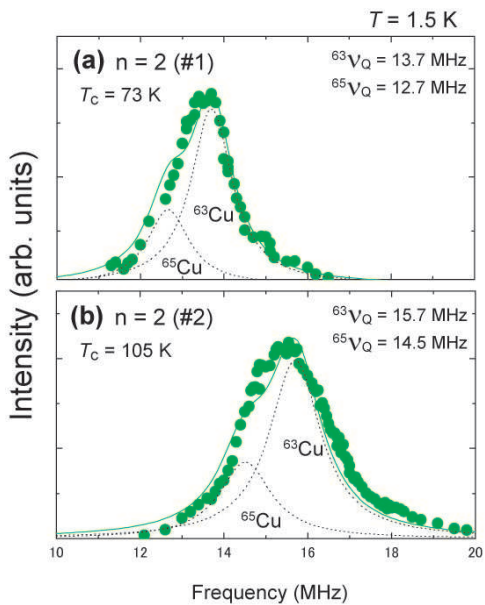}
\end{center}
\caption{\footnotesize (color online)  Cu-NQR spectra for the $n=2$ compounds (a): \#1 and (b): \#2, the respective $^{63}\nu_{\rm Q}$s of which were estimated to be 13.7 MHz and 15.7 MHz. The solid  curves indicate the simulated results arising from two NQR spectra from $^{63}$Cu and $^{65}$Cu. Here the linewidths for $n=2$ compounds(\#1) and (\#2) were 1.2 and 1.5 MHz, respectively. An increase of $^{63}\nu_{\rm Q}$ to 15.7 MHz in the $n=2$ compound(\#2) is due to the increase of the hole doping level.}
\label{fig:NQR}
\end{figure}

Figure \ref{fig:spe2F2} shows the $^{63}$Cu-NMR spectrum for the $n=2$ compound(\#1) at 20 K with $H$ perpendicular to the $c$-axis 
at a fixed frequency of 174.2 MHz. The largest peak at around $^{63}K=0$ arises from a central peak ($I=-1/2 \Leftrightarrow$ 1/2 transition) for $^{63}$Cu.  When we consider the second-order NQI for eq.(\ref{eq:hamiltonian}) under the condition of ${\cal H_{\rm Z}}$ $\gg$ ${\cal H_{\rm Q}}$, an NMR frequency $\omega_0^{2nd}$ at this central peak is given by
\begin{equation}
\frac{\omega_0^{\rm 2nd}-\gamma_{\rm N}H_{\rm res}}{\gamma_{\rm N}H_{\rm res}}=K_\perp+\frac{3\nu_{\rm Q}^2}{16(1+K_\perp)}\frac{1}{(\gamma_{\rm N}H_{\rm res})^2},
\label{eq:shift}
\end{equation}
where $H_{\rm res}$ is an NMR resonance field and $K_\perp$, the Knight shift with $H$ perpendicular to the $c$-axis. 
The inset of Fig.~\ref{fig:spe2F2} shows a plot of $(\omega_0^{2nd}-\gamma_{\rm N}H_{\rm res})/\gamma_{\rm N}H_{\rm res}$ vs. $(1/\gamma_{\rm N}H_{\rm res})^2$, which enables us to estimate  $^{63}\nu_{\rm Q}\sim$ 13~MHz from the slope in this plot.  
The NMR frequencies $\omega_{\rm 0}^{1st}$ for two satellite peaks ($\pm1/2 \Leftrightarrow \pm3/2$) due to the first-order NQI  are given by 
\begin{equation}
\frac{\omega_0^{\rm 1st}-\gamma_{\rm N }H_{\rm res}}{\gamma_{\rm N} H_{\rm res}}=K_{\perp} \mp \frac{\nu_{\rm Q}}{2}\frac{1}{\gamma_{\rm N} H_{\rm res}} .
\end{equation} 
The peak at $H=14.85$ T in Fig. \ref{fig:spe2F2} arises from the ($-$1/2 $\Leftrightarrow$ $-$3/2) transition of $^{63}$Cu, confirming $^{63}\nu_{\rm Q}\sim$ 13$-$14 MHz. However, the most accurate measurement of $^{63}\nu_{\rm Q}$ comes from the NQR spectrum. Figures \ref{fig:NQR}(a) and (b) show the Cu-NQR spectra for the $n=2$ compounds(\#1) and (\#2), indicating $^{63}\nu_{\rm Q}=13.7$ MHz and 15.7 MHz, respectively. These values are comparable to $^{63}\nu_{\rm Q}=$ 15$-$17 MHz for hole-doped HgBa$_2$CaCu$_2$O$_y$ (Hg-1212), which includes two OPs in a five-fold pyramidal coordination of oxygen \cite{Ohsugi_JLTP}.
Note that these $^{63}\nu_{\rm Q}$ values for hole-doped OP are much larger than those of electron-doped  copper oxides. For example, $^{63}\nu_{\rm Q}$ for electron-doped Nd$_{1.85}$Ce$_{0.15}$CuO$_{4}$ is reduced to $^{63}\nu_{\rm Q}\le$0.5 MHz, smaller than $^{63}\nu_{\rm Q}$ = 14 MHz of the non-doped Nd$_{2}$CuO$_{4}$ by more than an order of magnitude. This is because electron-doping into the Cu-$3d$ orbit decreases the on-site electronic term in the EFG; the on-site term is the main contribution to $\nu_{\rm Q}$ at Cu sites for high-$T_{\rm c}$ compounds \cite{Zheng,e_nuQ1,e_nuQ2,e_nuQ3,e_nuQ4,e_nuQ5}.

\begin{figure}[htpb]
\begin{center}
\includegraphics[width=0.85\linewidth]{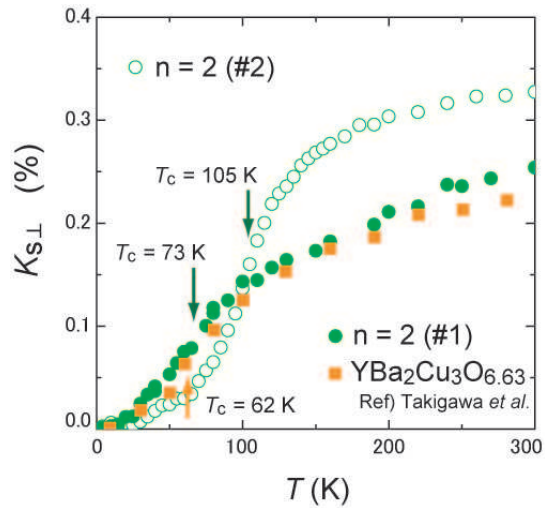}
\end{center}
\caption{\footnotesize (color online)  The $T$ dependencies of the Knight shift ($K_{\rm s\perp}(T)$) for the $n=2$ compounds(\#1) and (\#2). $K_{\rm s\perp}(T)$ for the $n=2$ compound(\#1) is quite similar to that for underdoped YBa$_2$Cu$_3$O$_{6.63}$ ($T_{\rm c}=62$ K) \cite{Takigawa}. 
$K_{\rm s\perp}(T)$ in the normal state decreases upon cooling, reflecting an opening of pseudogap, which is characteristic for the underdoped high-$T_{\rm c}$ compounds with hole carriers. 
By using the experimental relation between $K_{\rm s\perp}(T)$ at room temperature and a doping level at each layer\cite{Zheng, Tokunaga_JLTP, Kotegawa2001}, $N_{\rm h}\sim $0.174 and $N_{\rm h}\sim $0.225 are estimated for the $n=2$ compounds(\#1) and (\#2), respectively.}
\label{fig:Knight2}
\end{figure}

The Knight shift ($K$) is determined from the central peak ($-$1/2 $\Leftrightarrow$ $+$1/2) in the $^{63}$Cu-NMR spectrum (see Fig. \ref{fig:b2}(a)) by subtracting the second-order shift of NQI according to eq.(\ref{eq:shift}). In general, $K$ comprises the $T$-dependent spin part ($K_{\rm s}(T)$) and the $T$-independent orbital part ($K_{\rm orb}$); $K=K_{\rm s}(T)+K_{\rm orb}$. Here, $K_{\rm orb}$ was determined as 0.23 ($\pm0.02$)\%, assuming $K_{s}\approx 0$ in the $T=0$ limit. Note that $K_{\rm orb}$ is not so different among high-$T_{\rm c}$ compounds regardless of either IP or OP (as example see refs. \cite{Magishi, Julien, Barrett}).  Figure \ref{fig:Knight2} shows $K_{\rm s\perp}(T)$ for the $n=2$ compounds(\#1) and (\#2) with $H$ perpendicular to the $c$-axis, along with the data for YBa$_2$Cu$_3$O$_{6.63}$ (YBCO$_{6.63}$) (from ref.\cite{Takigawa}). $K_{\rm s\perp}(T)s$ for these samples decrease upon cooling, followed by a steep decrease below $T_{\rm c}$. These behaviors are typical of the high-$T_{\rm c}$ copper oxides with hole carriers, reflecting the opening of a pseudo gap \cite{Kotegawa2001,RE,REbook,Takigawa}. Here, note that the behavior of $K_{\rm s\perp}(T)$ in hole-doped cuprates is totally different from that in electron-doped ones such as Pr$_{0.91}$LaCe$_{0.09}$CuO$_{4-y} $ \cite{Zheng_edope}. Using the empirical relation between $K_{\rm s\perp}(T)$ at room temperature and the local hole density ($N_{\rm h}$) in a CuO$_2$ plane \cite{Zheng, Tokunaga_JLTP, Kotegawa2001}, $N_{\rm h}$ in the $n=2$ compound(\#1) with $T_{\rm c}=73$ K is estimated to be $N_{\rm h}\sim$ 0.174, which is comparable to $N_{\rm h}\sim$ 0.16 in YBCO$_{6.63}$ with $T_{\rm c}=62$ K.  
The $N_{\rm h}$ of the $n=2$ compound(\#2) is also estimated to be $\sim 0.225$, larger than that of the $n=2$ compound(\#1). The increase of hole-doping level increases $^{63}\nu_{\rm Q}$ from 13.7 MHz to 15.7 MHz, which is a character of hole-doped copper oxides \cite{Ohsugi, Ohsugi_JLTP, Zheng}.

Although the $n=2$ compound(\#1) was expected to be in a Mott insulating state in the ideal nominal composition, it was found that the hole carriers of $N_{\rm h}\sim$ 0.174 are unexpectedly introduced.
We consider that the hole doping is probably ascribed to the unexpected substitution of O$^{-2}$ for F$^{-1}$ at the apical site. 
At this stage, we would imply that an effect of off-stoichiometry results in doping carriers in Ba$_2$Ca$_{n-1}$Cu$_n$O$_{2n}$F$_2$.

\subsection{$^{63}$Cu-NMR studies on the $n$=3 and $n$=4 compounds}

An important finding here is that the $n=2$ compound(\#1) is a hole-doped high-$T_{\rm c}$ superconductor; F$^{-1}$ at the apical site is possibly substituted by O$^{-2}$. This result raises a question as to whether the {\it self-doping} mechanism actually occurs in the $n=3$ and $n=4$ compounds. The {\it self-doping} mechanism was supported by a previous study \cite{Shimizu} on the assumption of the nominal composition, when the Cu valence on average was +2 due to the total charge neutrality. It is important to reconsider this earlier conclusion in light of the present results.
%
\begin{figure}[tpb]
\begin{center}
\includegraphics[width=0.75\linewidth]{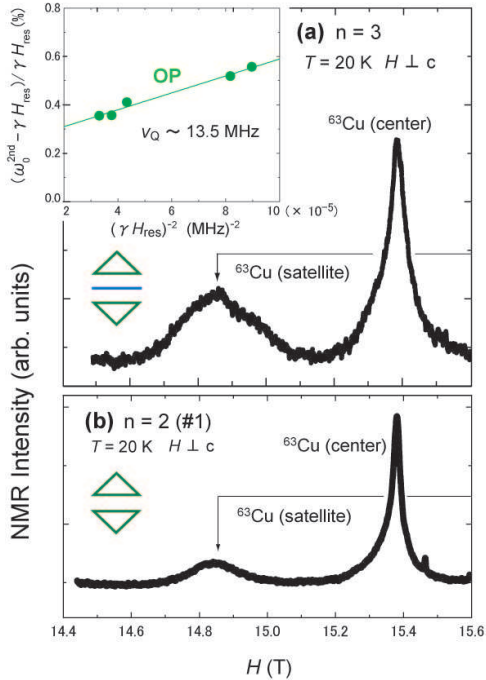}
\end{center}
\caption{\footnotesize (color online) The $^{63}$Cu-NMR spectra for (a) the $n$ = 3 compound and (b) the $n=2$ compound(\#1) (shown again) at 20 K with $H$ perpendicular to the $c$-axis. In (a), the central and the satellite peaks arise from the OP; whereas, the NMR spectrum for the IP is not observed because of the development of strong magnetic correlations. From the plot in the inset of (a), $^{63}\nu_{\rm Q}$ for the OP is evaluated to be $\sim$ 13.5 MHz (see the text), which is almost the same as that for the OP in the $n=2$ compound(\#1). }
\label{fig:b1}
\end{figure}
%
\begin{figure}[h]
\begin{center}
\includegraphics[width=0.75\linewidth]{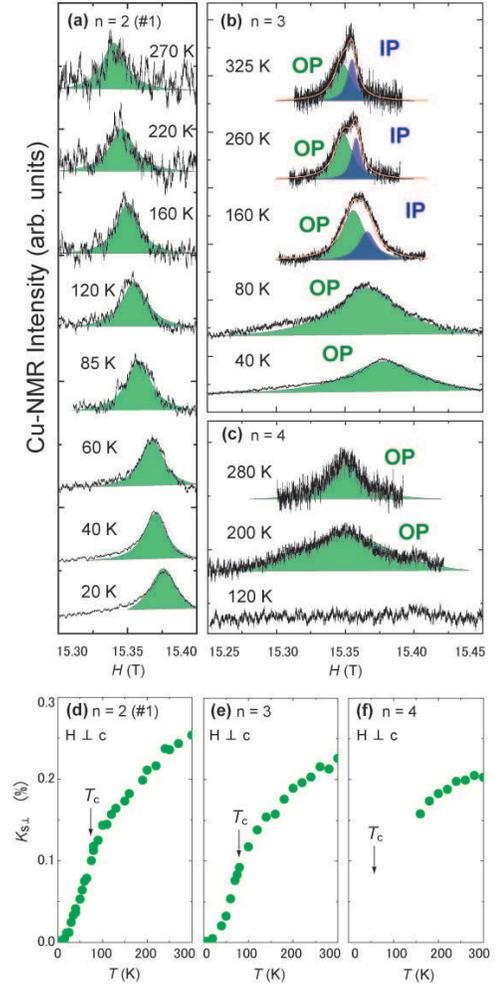}
\end{center}
\caption{\footnotesize (color online)  (a-c) The $T$ dependences of $^{63}$Cu-NMR spectra for the $n$ = 2 (\#1), $n$ = 3, and $n$ = 4 compounds. In (b), the asymmetric spectra for the $n$ = 3 compound above 160 K can be reasonably explained by the two spectra arising from the IP and the OP. Below 160 K, the Cu-NMR spectrum from the IP disappears  due to short relaxation times $1/T_2$ and $1/T_1$ in association with a possible development of strong magnetic correlations upon cooling. In (c), the $^{63}$Cu-NMR spectrum was observed only above 160 K, arising from the OP, but not from the IP. This OP-NMR spectrum disappears at low temperatures, implying that a spontaneous moment due to the AFM order develops even at the OP, followed by the onset of SC  below $T_{\rm c}$ = 55 K.
(d-f) The $T$-dependence of $K_{\rm s\perp}(T)$ for the $n$ = 2 (\#1), $n$ = 3, and $n$=4 compounds. Note that $K_{\rm s\perp}(T)$s for the $n$ = 2 (\#1) and $n$ = 3 compounds decrease rapidly below $T_{\rm c}$. }
\label{fig:b2}
\end{figure}

\begin{figure}[htbp]
\begin{center}
\includegraphics[width=0.7\linewidth]{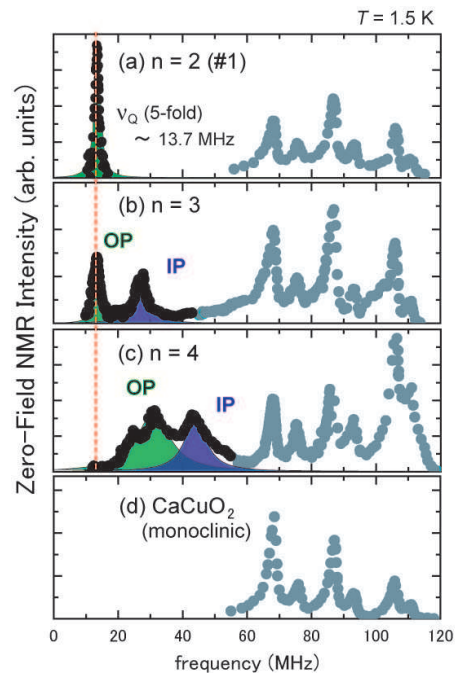}
\end{center}
\caption{\footnotesize (color online) Cu-ZFNMR/NQR spectra at $T=$ 1.5 K. (a)~The Cu-NQR spectrum of the OP for the $n$ = 2 compound(\#1) with $^{63}\nu_{\rm Q}=$ 13.7 MHz. (b)~The Cu-NQR/ZFNMR spectra for the $n$ = 3 compound. The NQR spectrum for the OP is observed to have almost the same value of $^{63}\nu_{\rm Q}$ as that of the OP for the $n$ = 2 compound(\#1), revealing that no static ordered moment is induced at low temperatures.  The spectrum observed around 27 MHz arises from the IP with $H_{\rm int}\sim$ 2.4 T and hence an AFM moment $M_{\rm AFM}\sim0.12$ $\mu_{\rm B}$ (discussed in detail in text). (c) The Cu-ZFNMR spectrum for the $n$ = 4 compound with $H_{\rm int}\sim$ 2.7 T and 3.8 T for the OP and the IP, respectively. Here, $M_{\rm AFM}=0.11$ $\mu_{\rm B}$ and 0.18 $\mu_{\rm B}$ are estimated for the OP and IP, respectively.  (d) The Cu-ZFNMR spectra for the monoclinic phase of CaCuO$_2$(II) obtained by the high-pressure synthesis method \cite{Takano}. The spectra observed in $f=$50-120 MHz in (a-c) arise from the impurity phase only via the synthesis condition at a high pressure and temperature (see text). In the previous paper \cite{Shimizu}, these spectra were wrongly believed to arise from the electron-doped OP.
}
\label{fig:zero}
\end{figure}

Figures \ref{fig:b1}(a) and (b) show the $^{63}$Cu-NMR spectra for the OP of the $n$ = 3 compound and the $n$ = 2 compound(\#1), respectively. The narrow peak in Fig. \ref{fig:b1}(a) arises from the ($-$1/2 $\Leftrightarrow$ 1/2) transition. The inset of Fig. \ref{fig:b1}(a) shows a plot of $(\omega_0^{2nd}-\gamma_{\rm N}H_{\rm res})/\gamma_{\rm N}H_{\rm res}$ as a function of $(1/\gamma_{\rm N}H_{\rm res})^2$, which enables us to estimate $^{63}\nu_{\rm Q}\sim$ 13.5~MHz for the OP in the $n$ = 3 compound. This value is consistent with the Cu-NQR measurement, which shows $^{63}\nu_{\rm Q}\sim$13.5 MHz in Fig. \ref{fig:zero}(b), and also consistent with the value confirmed from a satellite peak of the ($-$1/2 $\Leftrightarrow$ $-$3/2) transition due to the first-order NQI. In the previous paper (see Fig.3 in ref.\cite{Shimizu}), the narrow and satellite peaks in Fig. \ref{fig:b1}(a) were wrongly assigned to the ($-$1/2 $\Leftrightarrow$ 1/2) transitions for the IP and the OP, respectively. 
Meanwhile, we have succeeded in observing the asymmetric $^{63}$Cu-NMR spectrum at high temperatures above 160 K for the $n$ = 3 compound, as shown in Fig. \ref{fig:b2}(b), which can be reasonably decomposed into two lines arising from the IP and the OP.
Below 160 K, the Cu-NMR spectrum from the IP disappears due to the short relaxation rates $1/T_2$ and $1/T_1$ in association with the development of strong magnetic correlations toward an AFM transition upon cooling. Such $T$ dependence of Cu-NMR spectra were also observed in the five-layered compounds Hg-1245 and Tl-1245 \cite{Kotegawa2004}; in the two compounds, the spectra of three IPs disappear upon cooling due to the onset of AFM order, and those of the OP become broader as in the $n$ = 3 compound. 
As presented in section {\bf D}, such behavior in the $n$ = 3 compound is relevant with an onset of the AFM order below $T_{\rm N}\sim$ 23 K. 
The similarities among the $n =$ 3 compound, Hg-1245, and Tl-1245 ensure that the $n$ = 3 compound is hole-doped, and also that the hole-doped IP shows AFM transition at low temperature.  In fact, as discussed in the next section, the observation of Cu-NMR at $H$=0 is a direct indication of static AFM order in the $n$ = 3 compound.

In the $n$ = 4 compound, the $^{63}$Cu-NMR spectrum was observed only above 160 K as shown in Fig. \ref{fig:b2}(c); the spectrum arises only from the OP, but not from the IP. Unexpectedly, this OP-NMR spectrum disappears at low temperatures, suggesting that a spontaneous moment due to an AFM order develops even in the OP, followed by the onset of SC below $T_{\rm c}$ = 55 K. 
We consider that the $n=4$ compound is also hole-doped with less hole carriers than Hg-1245 and Tl-1245, so that even the OPs undergo the AFM order, as discussed later in detail.
Figures \ref{fig:b2}(e) and (f) respectively indicate $K_{\rm s\perp}(T)$s for the OPs in the $n$ = 3 and $n$ = 4 compounds. From $K_{\rm s\perp}(T)$s at room temperature, $N_{\rm h}\sim$~0.160 and $N_{\rm h}\sim$~0.148 are estimated at the respective OPs in the $n$=3 and $n$=4 compounds, confirming that the $n=4$ compound is actually heavily underdoped. 
Together, all these data suggest that the OP and the IP in these compounds are doped by hole carriers. In this context, the {\it self-doping} mechanism does not occur in the multilayered copper oxides with apical fluorines.

\subsection{Zero-field Cu-NMR and Cu-NQR studies on the $n$ = 2, $n$ = 3, and $n$ = 4 compounds}

The observation of a zero-field NMR (ZFNMR) spectrum enables us to assure an onset of magnetic order, since magnetically ordered moments induce $H_{\rm int}$ at nuclear sites. Here, the nuclear Hamiltonian given by eq.(\ref{eq:hamiltonian}) is described with $H_{\rm int}$ instead of $H$. Figure \ref{fig:zero}(a) shows the Cu-NQR spectrum for the $n=$2 compound(\#1) at 1.5 K (shown again), along with the Cu-ZFNMR/NQR spectra for (b) the $n=$3 and (c) $n=$4 compounds.
In the previous study \cite{Shimizu}, the spectra in the range of 50$-$120 MHz for the $n$ = 3 and $n$ = 4 compounds were assigned as arising from the OPs.  However, we studies multilayered compounds prepared in various synthesis conditions under pressure, and then revealed that these lines arise from an impurity phase, namely monoclinic CaCuO$_2$(II) \cite{Takano}. The ZFNMR spectrum of CaCuO$_2$(II) is shown in Fig.\ref{fig:zero}(d). The samples used in this paper were synthesized using high-pressure technique such as $P=$ 4.5 GPa and $T=$ 1050 $^\circ$C. In order to confirm that the spectra in the range of 50$-$120 MHz originate from CaCuO$_2$(II), we synthesized Hg-based five-layered compounds by two different synthesis conditions, $P=$ 4.5 GPa and $T=$ 1050 $^\circ$C, and $P=$ 2.5 GPa and $T=$ 950 $^\circ$C. Actually, it was confirmed that the extrinsic impurity phase of CaCuO$_2$(II) was included in the sample at the former condition, but not at the latter one \cite{Mukuda2008}. Consequently, the NMR spectra in the range below $\sim$50 MHz are anticipated to be intrinsic, arising from the OP and the IP. The presence of CaCuO$_2$(II) was not detected by XRD measurement, but ZFNMR. Here, note that it is very difficult to determine the amount of CaCuO$_2$(II) as impurity from the ZFNMR intensity because the experimental condition is not always the same in low and high frequency ranges, in particular when ZFNMR spectra are widely distributed over a wide frequency range just as in the present case.
We emphasize here that the SC and magnetic properties inherent to the intrinsic phase does not depend on the synthesis condition.
We note that the presence of such impurity domains may be one of possible reasons for off-stoichiometry existing in these compounds. 

Figure \ref{fig:zero}(a) indicates the Cu-NQR spectrum for the $n$ = 2 compound(\#1). 
As discussed in section {\bf A}, a single sharp peak at 13.7 MHz corresponds to $^{63}\nu_{\rm Q}$ for the OP, indicating that its ground state is non-magnetic with $T_{\rm c}=73$ K and $N_{\rm h}\sim$ 0.174.  Figure \ref{fig:zero}(b) shows the Cu-ZFNMR/NQR spectra for the $n$ = 3 compound. Two peaks are observed separately; one at 13.5 MHz, which is almost the same frequency as the OP in the $n$ = 2 compound(\#1), consistent with the value estimated from the second-order shift due to NQI (see the inset of Fig. \ref{fig:b1}(a)). Accordingly, another peak is assigned to the IP. An NQR frequency at IP is reported as $^{63}\nu_{\rm Q}\sim$ 8$-$9 MHz for other multilayered systems \cite{Tokunaga,Kotegawa2001,Kotegawa2004}, which is significantly smaller than OP. 
This is because the oxygen coordination of the Cu site at IP is smaller than that at OP; OP has a five-fold pyramidal coordination, while IP has a four-fold square coordination (see Figs. \ref{fig:crystal}(a-c)).
Using $^{63}\nu_{\rm Q}=8 (\pm 1)$ MHz, $H_{\rm int}$ is estimated to be $\sim$2.4 T for the IP. $H_{\rm int}$ at the CuO$_2$ plane is generally given by $H_{\rm int}=|A_{\rm hf}|M_{\rm AFM}=|A-4B|M_{\rm AFM}$, where $A$ and $B$ are the on-site hyperfine field and the supertransferred hyperfine field from the four nearest neighboring Cu-AFM moments, respectively, and $M_{\rm AFM}$ is the AFM moment \cite{MilaRice}. Here $A\sim$ 3.7 T/$\mu_{\rm B}$, $B({\rm OP})\sim$ 7.4 T/$\mu_{\rm B}$, and $B({\rm IP})\sim$ 6.1 T/$\mu_{\rm B}$ are assumed to be the same as those for Hg-1245 \cite{Kotegawa2004}. Using these values, a uniform AFM moment at the IP is estimated as $M_{\rm AFM}$(IP) $\sim$ 0.12 $\mu_{\rm B}$ in the $n$ = 3 compound.

Figure \ref{fig:zero}(c) shows the ZFNMR spectrum for the $n$ = 4 compound, which exhibits two peaks at around 30 and 45 MHz. No NQR signals are observed around $^{63}\nu_{\rm Q}$(IP) $\sim$ 8$-$9 MHz and $^{63}\nu_{\rm Q}$(OP) $\sim$ 13.5 MHz, which are expected for a non-magnetic state. 
The respective $M_{\rm AFM}=0.11$ $\mu_{\rm B}$ and 0.18 $\mu_{\rm B}$ for the OP and the IP are estimated from $H_{\rm int}$ = 2.7 T and 3.8 T corresponding to the ZFNMR frequencies of 30 MHz and 45 MHz. $M_{\rm AFM}$(IP) is larger than $M_{\rm AFM}$(OP)  because $N_{\rm h}$(IP) is lower than $N_{\rm h}$(OP) = 0.148 due to the charge imbalance between the OP and the IP in multilayered systems in general. 
We should note that the OP is mainly responsible for the SC with $T_{\rm c}$ = 55 K, since $N_{\rm h}$(OP)$=0.148 > N_{\rm h}$(IP). 
Therefore, we revealed that the uniform mixing of AFM  with $M_{\rm AFM}=0.11$ $\mu_{\rm B}$ and SC occurs in the OP, reminiscent of the uniform mixing of AFM ($M_{\rm AFM}=0.1$ $\mu_{\rm B}$ and $T_{\rm N}=55$ K) and SC ($T_{\rm c}$ = 85 K) in the three IPs in the five-layered Hg-1245 system \cite{Mukuda2008}. 

\begin{figure}[t]
\begin{center}
\includegraphics[width=0.8\linewidth]{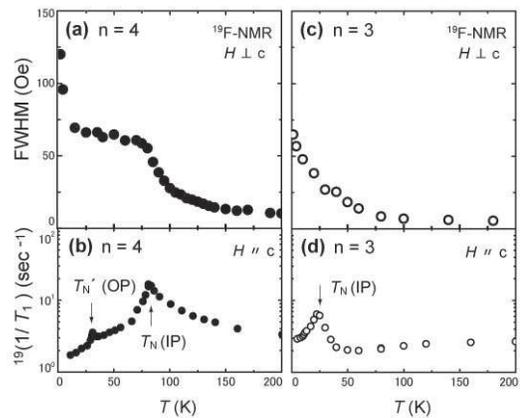}
\end{center}
\caption{\footnotesize (color online) $T$ dependences of a full-width at a half maximum (FWHM) in $^{19}$F-NMR spectrum  for (a) the $n$ = 4  and (c) $n$ = 3 compounds with $H \sim 4.35$ T perpendicular to the $c$-axis. The FWHM for the $n$ = 4 compound increases upon cooling below $\sim$ 100 K and tends to saturate below $\sim$ 75 K. This behavior resembles the increasing behavior of $M_{\rm AFM}$ following 
the AFM order parameter, demonstrating that the AFM ordering occurs in the $n=$ 4 compound.
In the $n=$ 3 compound, the FWHM also increases affected by the AFM order. 
The $T$ dependences of $^{19}(1/T_1)$ for (b) the $n$ = 4 and (d) $n$ = 3 compounds at $f=$ 174.2 MHz with $H \sim 4.35$ T parallel to the $c$-axis. $^{19}(1/T_1)$ is enhanced by critical magnetic fluctuations developing toward a magnetically ordered temperature and has a peak at $T_{\rm N}$ = 80 K and 23 K for the $n$ = 4 and 3 compounds, respectively. Note that a small peak in $^{19}(1/T_1)$ is also observed around 30 K. This low-$T$ peak seems to be relevant with $T_N'$ = 30 K inherent to the OP, below which $M_{\rm AFM}$(OP) becomes completely static and develops upon cooling, increasing the $^{19}$FWHM rapidly at temperatures lower than 30 K (see text).}
\label{fig:F-NMR}
\end{figure}

\subsection{Magnetic ordering temperature in the $n$ = 3 and $n$ = 4 compounds}

A $^{19}$F-NMR spectrum and the nuclear-spin-lattice relaxation rate $^{19}(1/T_1)$ have been measured to determine the N\'eel temperature ($T_{\rm N}$) in these compounds. Figures \ref{fig:F-NMR}(a) and (b) show the $T$ dependences of the $^{19}$F-NMR spectral width, defined as a full-width at a half maximum (FWHM), and $^{19}(1/T_1)$ for the $n$ = 4 compound. Here, $H$ is applied perpendicular and parallel to the $c$-axis. As reported in the previous paper \cite{Shimizu}, the FWHM for the $n$ = 4 compound increases upon cooling below $\sim$100 K and tends to saturate below $\sim$75 K. This resembles the increasing behavior of $M_{\rm AFM}$ following the AFM order parameter, demonstrating that the AFM ordering occurs in the $n$ = 4 compound. This is corroborated by the observation of a peak in $^{19}(1/T_1)$ around 80 K, as shown in Fig. \ref{fig:F-NMR}(b). This peak originates in the slowing down of critical magnetic fluctuations as the magnetic ordering temperature is approached. We suppose that the gradual increase in FWHM upon cooling below 150 K may be ascribed to the development of short-range AFM order in association with the distribution of $T_{\rm N}$, although its origin is not yet known.

As for the $n$ = 3 compound, the $T$ dependences of the FWHM for the $^{19}$F-NMR, and $^{19}(1/T_1)$ 
 are shown in Figs. \ref{fig:F-NMR}(c) and (d). Remarkably, the peak in $^{19}(1/T_1)$ is observed at 23 K, as shown in Fig. \ref{fig:F-NMR}(d). Along with the observation of the spontaneous AFM moment $M_{\rm AFM}=0.12$ $\mu_{\rm B}$ at the IP at $T=$ 1.5 K, this suggests the onset of AFM order for the $n$ = 3 compound as well as the $n$ = 4 compound. The AFM order is triggered by the single IP with $M_{\rm AFM}=0.12$ $\mu_{\rm B}$ through the interlayer magnetic coupling ($J_{\rm inter}$) via the three layers, namely, SC-OP, BaF, and SC-OP (see Fig. \ref{fig:crystal}(b)). 

Interestingly, as for the $n$ = 4 compound, a small peak in $^{19}(1/T_1)$ is also observed around 30 K, as indicated in Fig. \ref{fig:F-NMR}(b). Moreover, the $^{19}$FWHM rapidly increases upon cooling at low temperatures, as shown in Fig. \ref{fig:F-NMR}(a). We assume that this low-$T$ peak in $^{19}(1/T_1)$ originates in the slowing down of critical magnetic fluctuations at the OP, below which $M_{\rm AFM}$(OP) becomes completely static and develops upon cooling.  
Since the two underdoped IPs have $M_{\rm AFM}$(IP) $\sim$ 0.18 $\mu_{\rm B}$ larger than $M_{\rm AFM}$(OP) $\sim 0.11$ $\mu_{\rm B}$, it is expected that the two IPs trigger the AFM order below $T_{\rm N}=$ 80 K by an interlayer magnetic coupling ($J_{\rm inter}$) through the three layers, namely, SC-OP, BaF, and SC-OP. On the other hand, the OPs trigger the SC with $T_{\rm c}=55$ K because $N_{\rm h}$(OP) is larger than $N_{\rm h}$(IP). Since $M_{\rm AFM}$(OP) is smaller than $M_{\rm AFM}$(IP) and the interlayer magnetic coupling $J_{\rm inter}$ is significantly smaller than the in-plane superexchange coupling ($J_{\rm intra}$) in the CuO$_2$ plane, the N\'eel temperature inherent to the OP, $T_{\rm N}'$ = 30 K, seems to be smaller than that inherent to the I, $T_{\rm N}$ = 80 K. In this context, a spontaneous moment $M_{\rm AFM}$(OP) is probably not completely static in between $T_{\rm N}'$ = 30 K and $T_{\rm N}=80$ K; however it develops markedly up to 0.11 $\mu_{\rm B}$ upon cooling below 30 K. As a result, both AFM and SC uniformly coexist in the OP for the $n$ = 4 compound.

\section{summary}

\begin{figure}[h]
\begin{center}
\includegraphics[width=0.75\linewidth]{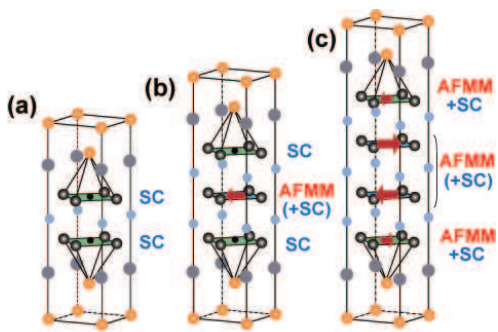}
\end{center}
\caption{\footnotesize (color online) An illustration of the ground-state properties for the (a) $n$ = 2 (\#1), (b) $n$ = 3, and (c) $n$ = 4 compounds. The outer planes (OPs) and the inner planes (IPs) are both doped by hole carriers and are in the underdoped regime with a hole doping level of $N_{\rm h}$(IP) $<$ $N_{\rm h}$(OP). (a) The $n$ = 2 compound(\#1) is the superconductor with $T_{\rm c}$ = 73 K in the underdoped regime at $N_{\rm h}$(OP) = 0.174, resembling the YBCO$_{6.63}$ with $T_{\rm c}=62$ K. (b) The $n=3$ compound is the antiferromagnetic superconductor with $T_{\rm c}$ = 76 K in which the underdoped IP with  $N_{\rm h}$(IP) $<$ $N_{\rm h}$(OP) = 0.160 is responsible for the AFM order with $M_{\rm AFM}$ = 0.12 $\mu_{\rm B}$ and $T_{\rm N}=23$ K.
(c) The $n=4$ compound is also the antiferromagnetic superconductor with $T_{\rm N}=80$ K higher than $T_{\rm c}=55$ K. The IPs trigger the AFM with $M_{\rm AFM}= 0.18$ $\mu_{\rm B}$ and $T_{\rm N}$ = 80 K, whereas the OPs trigger the SC with $T_{\rm c}=55$ K; interestingly, the latter shows the AFM order with $M_{\rm AFM}= 0.11$ $\mu_{\rm B}$ and also a possibility to have its own $T_{\rm N}'$ = 30 K, below which $M_{\rm AFM}$(OP) becomes completely static and develops upon cooling.}
\label{fig:summary}
\end{figure}

We have reported extensive studies by means of $^{63}$Cu-NMR/NQR and $^{19}$F-NMR on multilayered copper oxides Ba$_2$Ca$_{n-1}$Cu$_n$O$_{2n}$F$_2$ [denoted as the $n$ compound] with $n=2,3,4$. It is demonstrated that both the outer planes (OPs) and the inner planes (IPs) are doped by hole carriers with the doping level $N_{\rm h}$ of the OP being larger than that of the IP. As a result, we suggest the absence of the {\it self-doping} mechanism; the OP(IP) is electron-doped, whereas the IP(OP) is hole-doped, which was proposed in previous studies \cite{Shimizu,Chen,OK}, assuming a nominal content of fluorine at the apical site. 

The ground-state properties for the $n$ = 2 (\#1), $n$ = 3, and $n$ = 4 compounds are summarized in Fig.~\ref{fig:summary}. 
(i) The $n$ = 2 compound(\#1) is the superconductor with $T_{\rm c}$ = 73 K in the underdoped regime with $N_{\rm h}$(OP) = 0.174, resembling YBCO$_{6.63}$ with $T_{\rm c}=62$ K. (ii) The $n$ = 3 compound is the antiferromagnetic superconductor with $T_{\rm c}$ = 76 K in which the single IP is in the underdoped regime with $N_{\rm h}$(IP) $<$ $N_{\rm h}$(OP) = 0.160. $M_{\rm AFM}$ = 0.12 $\mu_{\rm B}$ triggers the AFM order at $T_{\rm N}=$23 K by the interlayer coupling via the three layers, namely, SC-OP, BaF, and SC-OP. (iii) The $n$ = 4 compound is also the antiferromagnetic superconductor with $T_{\rm N}=80$ K higher than $T_{\rm c}=55$ K. The two IPs trigger the AFM with $M_{\rm AFM}= 0.18$ $\mu_{\rm B}$ and $T_{\rm N}$ = 80 K, whereas the OP triggers the SC with $T_{\rm c}=55$ K; interestingly, the latter shows the AFM order with $M_{\rm AFM}= 0.11$ $\mu_{\rm B}$ and also a possibility to have its own $T_{\rm N}'$ = 30 K below which $M_{\rm AFM}$(OP) becomes completely static and develops upon cooling. We note that the increase in the number of IPs from one to two results in the strengthening of the interlayer coupling; $T_{\rm N}$ increases as the interlayer coupling becomes stronger, although the doping levels in the $n=3$ and 4 compounds are comparable. Consequently, we conclude that the uniform mixing of AFM and SC is a general property inherent to a single CuO$_2$ plane in an underdoped regime for hole-doping. This conclusion incorporates the ARPES results on the n=4 compound \cite{Chen}; it was found that the two Fermi sheets of the IP and OP are observed, and that the SC gap opens at the IP and OP below $T_{\rm c}$ = 55 K.

\section*{Acknowledgement}

The authors are grateful to M. Mori, T. Tohyama, and H. Eisaki for their helpful discussions. This work was supported by a Grant-in-Aid for Specially Promoted Research (20001004) and Scientific Research No. 17105002, and in part by Global COE Program (Core Research and Engineering of Advanced Materials-Interdisciplinary Education Center for Materials Science), from the Ministry of Education, Culture, Sports, Science and Technology (MEXT), Japan. One of the authors (S.S.) is financially supported as a JSPS Research Fellow.


\clearpage

\end{document}